\documentclass{webofc}
\usepackage[varg]{txfonts}   

\begin{document}
\title{First study of the initial gluonic fluctuations using UPCs with ALICE}
%
%

\author{\firstname{Adam} \lastname{Matyja}\inst{1}\fnsep\thanks{\email{adam.tomasz.matyja@cern.ch}}
        \firstname{} \lastname{for the ALICE Collaboration}
}

\institute{Institute of Nuclear Physics Polish Academy of Sciences, Radzikowskiego 152, 31-342 Krak\'ow, Poland
          }

\abstract{%
  Incoherent and dissociative $\rm{J/\psi}$ photoproduction is sensitive to fluctuations of the gluonic structure of the target. Thus, the measurement of $\rm{J/\psi}$ photoproduction of the colliding hadron sheds light on the initial state of QCD and provides important constraints on the initial conditions used in hydrodynamical models of heavy-ion collisions.
  The first measurement of the transverse momentum dependence of both coherent and incoherent $\rm{J/\psi}$ photoproduction in ultraperipheral Pb--Pb collisions at midrapidity is presented. These new results provide, for the first time, a clear indication of subnucleonic fluctuations of the lead target.
  We also present the new measurement of
  dissociative $\rm{J/\psi}$ photoproduction cross section
as a function of energy,
  in p--Pb collisions at forward rapidity.
  Dissociative results do not show any indication of saturation and agree with previous data.
}
\maketitle
\section{Introduction}
\label{intro}
When ions
do not touch each other (the impact parameter $b$ is larger than the sum of their radii) a collision is ultraperipheral (UPC). The moving ion emits photons.
The photon flux coming from one ion is enhanced by factor $Z^{2}$ ($Z=82$ for Pb) in comparison to proton ($Z=1$).
It induces large photon interaction cross section. Together with a low detector activity UPCs are ideal to study photoproduction of vector mesons (VMs), like $\rm{J/\psi}$. A virtual photon emitted by an ion can fluctuate into quark--antiquark pair which interacts with a target nucleus (coherent interaction) or nucleon inside nucleus (incoherent interaction) via the exchange of a color neutral particle, and the resulting VM is formed.
That is why the photoproduction cross section of $\rm{J/\psi}$ is particularly sensitive to the gluon density evolution at low Bjorken-$x$ ($x_{\rm{B}} = M_{\rm{VM}} \cdot \exp{(\pm y)} /\sqrt{s_{\rm{NN}}}$, where $M_{\rm{VM}}$ is a VM mass, $y$ is a VM rapidity, and $\sqrt{s_{\rm{NN}}}$ is centre-of-mass collision energy) at the leading order perturbative QCD calculations~\cite{Ryskin}. A hard scale, which allows for perturbative treatment, is assured by the high mass of $\rm{J/\psi}$ meson.
The higher photon-target centre-of-mass energy $W_{\gamma \rm{p, Pb}}=\sqrt{2E_{\rm{Pb}} M_{\rm{VM}} \exp{(\pm y)} }$, the lower $x_{\rm{B}}$ is accessible.
The photon-target energy or $x_{\rm{B}}$ dependence is sensitive to the gluon saturation effects in p--Pb collisions~\cite{MS_pPb,CCT} or both gluon shadowing~\cite{shadowing} and saturation~\cite{saturation} effects in Pb--Pb collisions~\cite{Escola_shadowing}. Additionally, a four-momentum transfer of target (Mandelstam $t$), $|t| \sim p_{\rm{T}}^2$, is sensitive to gluon distribution in the transverse plane, due to the fact that $b$ and $p_{\rm{T}}$ are Fourier conjugates. The $t$-dependent total cross-section can be calculated within Good--Walker (GW) approach~\cite{GoodWalker}.
The coherent part
informs about the average transverse gluon distribution.
The incoherent cross section
informs about variance of the spatial gluon distribution.
However, the GW approach has several limitations~\cite{Spencer}.

Theoretical calculations for both coherent and incoherent $\rm{J/\psi}$ photoproduction differential cross section have been carried out by several groups~\cite{CCT,MS_PbPb_coh_incoh,GKZ_coh,GKZ_inc,MSS}. Each group included in models quantum fluctuations either via initial event-by-event fluctuations in nucleon density profile or gluonic hot spots. By the definition, including quantum fluctuation has a stronger effect at larger $|t|$ values, because scattering happens on the smaller objects.
Similar mechanisms occur
for p--Pb collisions, where an ion emits a photon which probes the target proton. The interaction might be exclusive when photon couples to the whole proton or dissociative when photon couples to the proton inner structure and the proton breaks up.

The ALICE Collaboration has recently measured the $|t|$-dependent coherent~\cite{ALICE_coherent} and incoherent~\cite{ALICE_incoh} $\rm{J/\psi}$ cross section in UPC Pb--Pb at $\sqrt{s_{\rm NN}}=5.02$~TeV and energy dependent exclusive and dissociative $\rm{J/\psi}$ cross section in p--Pb collisions at  $\sqrt{s_{\rm NN}}=8.16$~TeV~\cite{ALICE_pPb}. These results are described below.
\section{$\rm{J/\psi}$ meson reconstruction in the ALICE detector}
\label{sec:ALICE}
Muons can be detected in the ALICE detector~\cite{ALICE_detector} at the pseudorapidity $-4<\eta<-2.5$ in the muon arm
or at the central region ($|\eta|<0.9$).
The $\rm{J/\psi}$ meson is reconstructed from two opposite charge muon tracks.
Any other activity in the detector is vetoed by the scintillator based V0 and AD forward detectors located at both sides and covering the pseudorapidity ranges $2.8<\eta<5.1$ and $-3.7<\eta<-1.7$ (V0), and $4.8<\eta<6.3$ and $-7.0<\eta<-4.9$ (AD). The  electromagnetic dissociation was checked by the neutron component of the Zero Degree Calorimeter (ZDC) covering $|\eta|>8.8$. The $\rm{J/\psi}$ yield was extracted by fitting muon-pair invariant mass distribution by an extended unbinned maximum likelihood fit describing a signal (single or double sided Crystal Ball) and a background (exponent) component. Various corrections coming from feed-down of $\psi'$, coherent, or incoherent production were applied based on $p_{\rm T}$ spectrum. The results were also corrected for the acceptance and reconstruction efficiency, pile-up, and electromagnetic dissociation due to veto from the forward detectors. In case of coherent production Bayesian unfolding was applied to account for $p_{\rm T}$ migrations and to transform from $p_{\rm T}^{2}$ to $|t|$.
The photon fluxes necessary for the cross section determination are taken from theory calculations.
\section{Coherent and incoherent $\rm{J/\psi}$ photoproduction.}
\label{sec:cohincohresults}
\begin{figure}[h]
  \centering
  \includegraphics[width=6.4cm,clip]{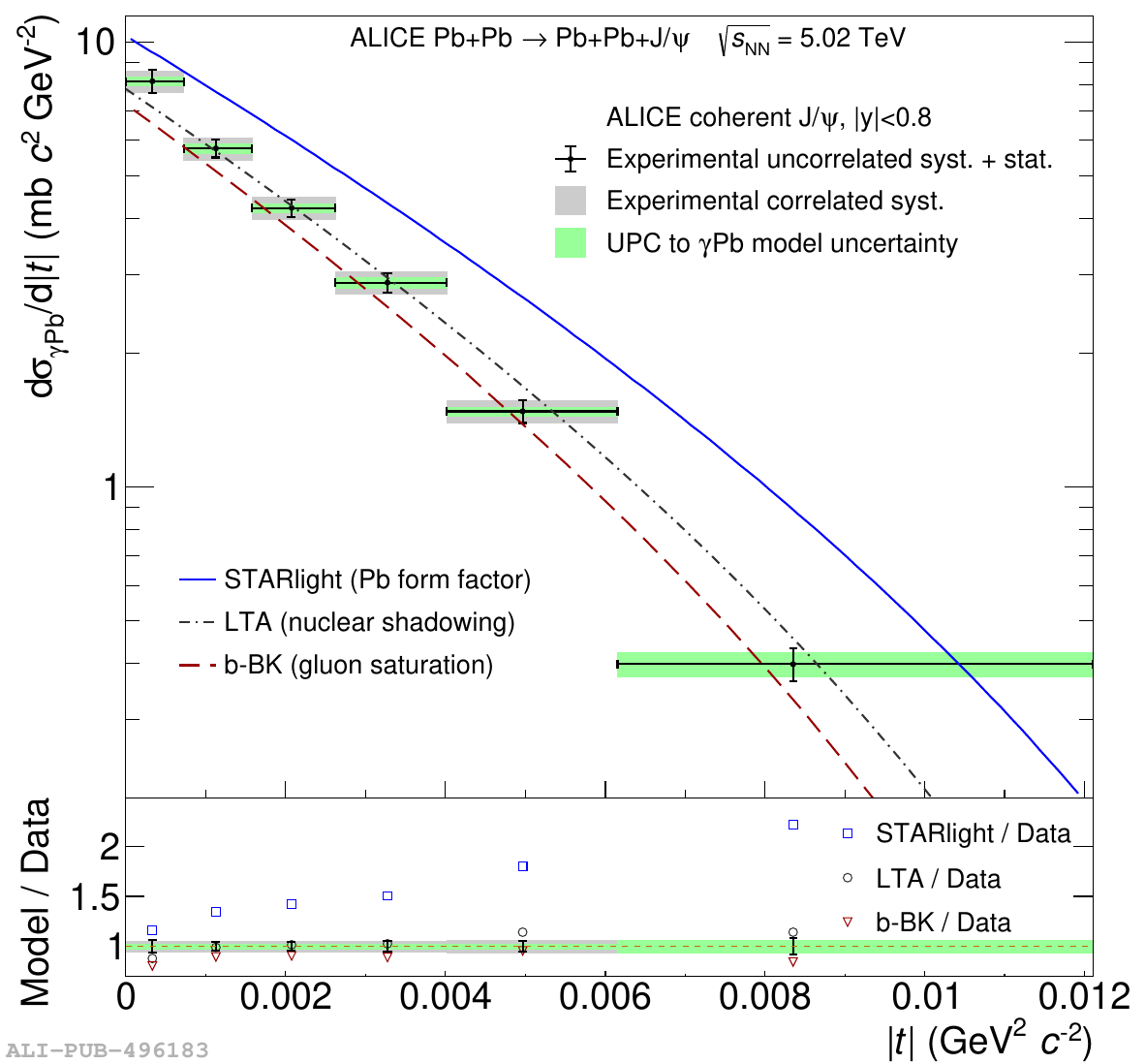}
  \includegraphics[width=6.4cm,clip]{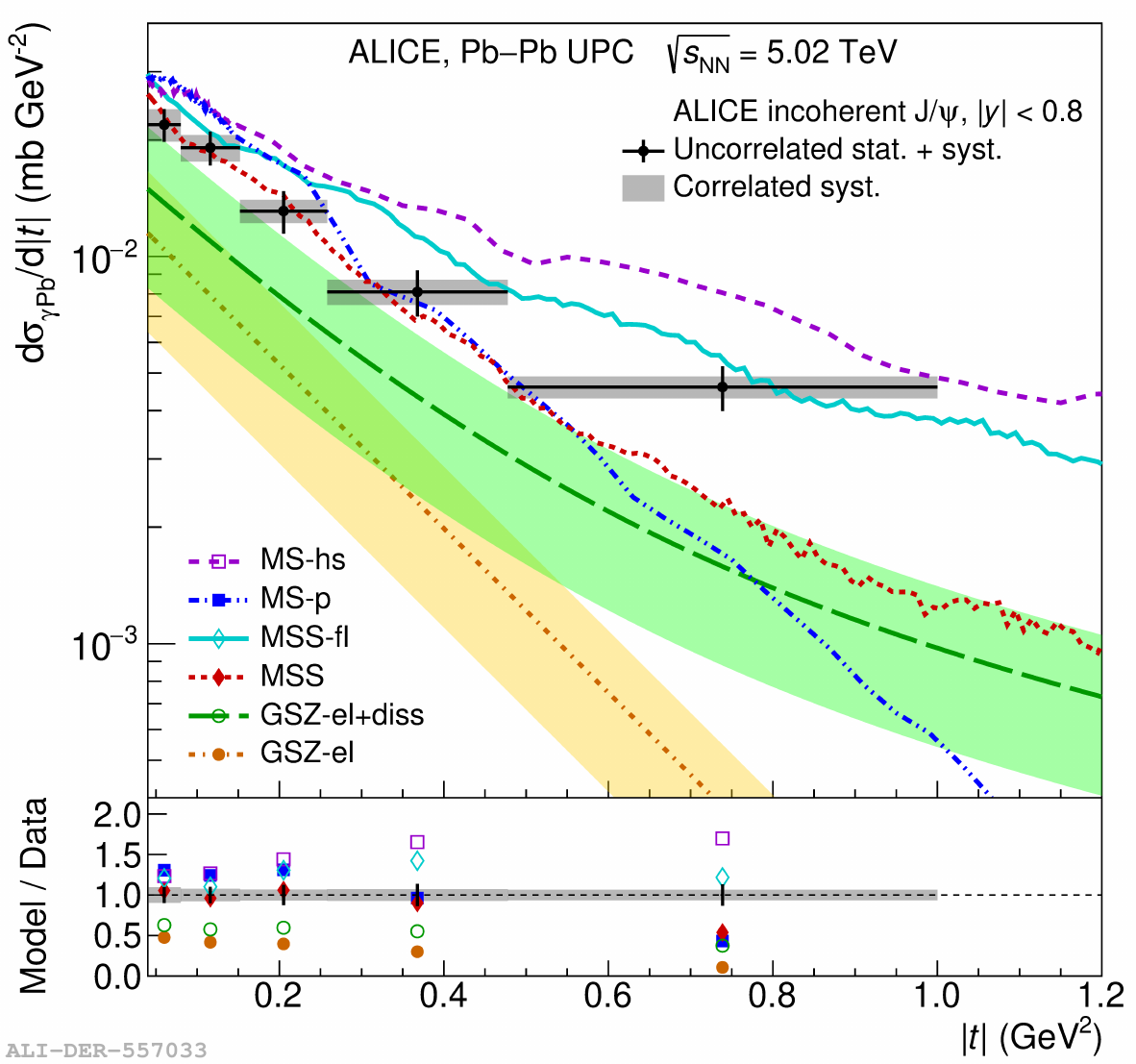}
  \caption{The $|t|$ dependence of the photonuclear cross section for the coherent (left) and the incoherent (right) photoproduction of $\rm{J/\psi}$ in UPC Pb--Pb at $\sqrt{s_{\rm NN}}=5.02$~TeV compared with theoretical models.}
  \label{fig:PbPb}       
\end{figure}

The $|t|$-dependent photonuclear cross section of coherently produced $\rm{J/\psi}$ is shown in the left panel of Fig.~\ref{fig:PbPb},
while for incoherently produced in the right panel. The coherent cross section results are compared to predictions of the STARlight Monte Carlo generator~\cite{starlight} and two models, LTA~\cite{GKZ_coh} based on nuclear shadowing and b-BK~\cite{bBK} based on gluon saturation. 
Both models reproduce the 
coherent $\rm{J/\psi}$ data rather well in contrary to STARlight which
overestimates the measured cross section. This disagreement might arise from the fact that STARlight
does not incorporate
shadowing.
The cross section of the incoherent photoproduction of $\rm{J/\psi}$ meson is compared with three theoretical predictions.
Two cases are considered for each calculation:
one including only elastic interaction with single nucleons, and another where the inner structure of nucleon is considered. 
The saturation model of M\"{a}ntysaari--Schenke (MS)~\cite{MS_PbPb_coh_incoh} 
provides two variants.
First variant (MS-hs) assumes that proton is composed of three hot spots whose positions in the impact parameter fluctuate event-by-event and there are fluctuations of the saturation scale. The subnucleon fluctuations are not considered in the second one (MS-p).
The first scenario of the perturbative GSZ model~\cite{GKZ_inc} includes only an elastic part of the incoherent cross section (GSZ-el). 
The other scenario includes additional part coming from dissociative component (GSZ-el+diss).
The last model, MSS~\cite{MSS} is also a saturation model 
based on JIMWLK equation. The inner structure of proton is included via hot spots.
The first scenario offers
initial density fluctuations (MSS-fl), and the other one does not include fluctuations (MSS).
Two aspects are considered when models are compared with data. There is a normalization, which is mainly linked to the scaling from proton to nuclear targets, and a slope driven by the size of the scattering object. None of the models describes both the normalization and the slope of the incoherent $\rm{J/\psi}$ data. Concerning normalization, the model should describe not only incoherent regime, but also the coherent one. The slope of the incoherent $\rm{J/\psi}$ data favors
models with subnucleon fluctuations,
which makes the distribution less steep.
\section{Dissociative $\rm{J/\psi}$ photoproduction}
\label{sec:dissociative}
\begin{figure}[h]
  \centering
    \includegraphics[width=6.4cm,clip]{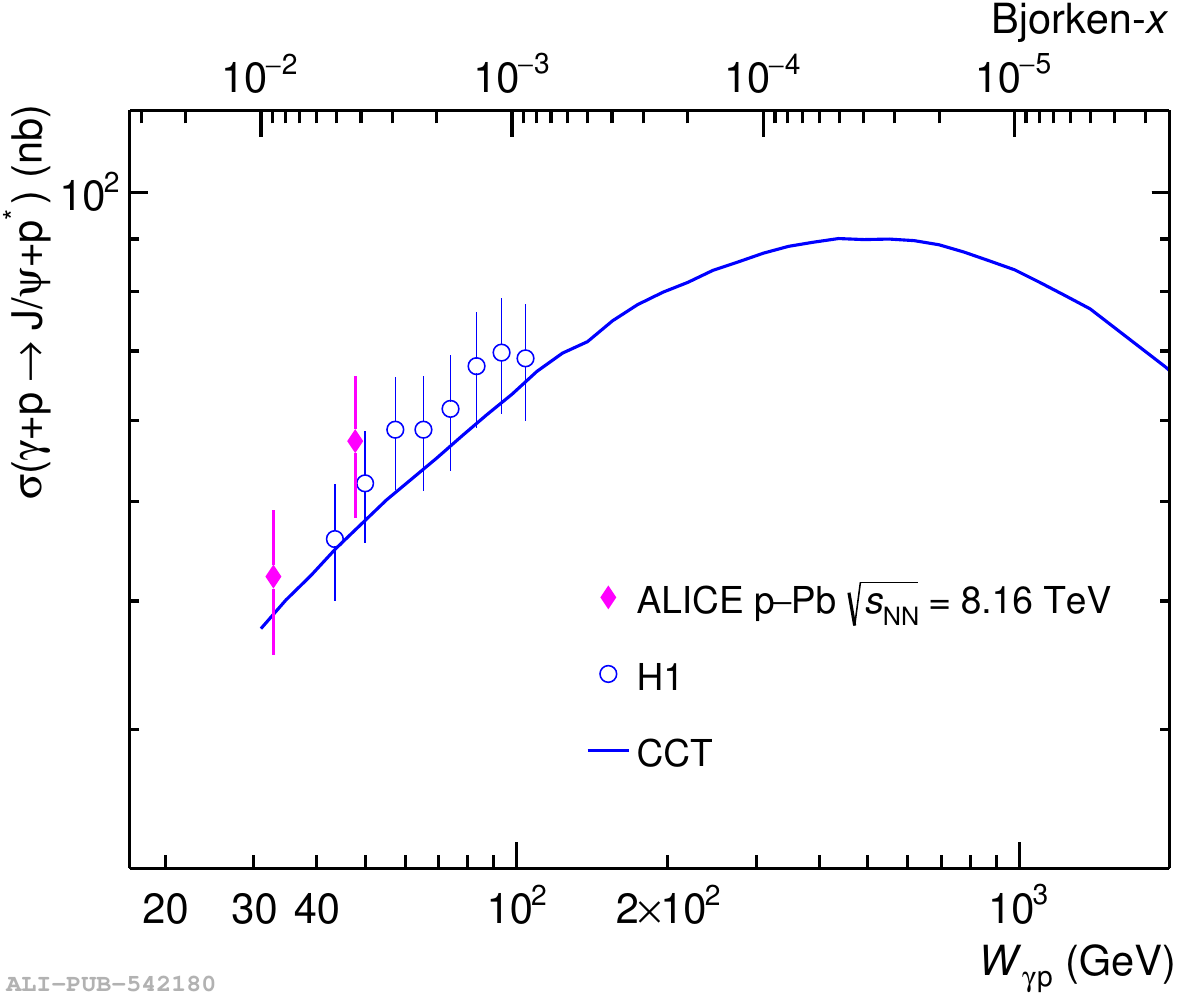}
  \caption{Dissociative $\rm{J/\psi}$ photoproduction cross section of protons measured by ALICE in p--Pb UPCs at $\sqrt{s_{\rm NN}} = 8.16$~TeV and compared with H1 data~\cite{H1}, and with the CCT model~\cite{CCT}.}
  \label{fig:pPb}       
\end{figure}

The dissociative $\rm{J/\psi}$ photoproduction cross section as a function of $W_{\rm \gamma p}$, covering the range $27<W_{\rm \gamma p}<57$~GeV is shown in Fig.~\ref{fig:pPb}. The ALICE measurement is compared with H1 result~\cite{H1}. There is a good agreement between the measurements. The experimental results are compared with the CCT model~\cite{CCT}.
The model considers energy dependent fluctuating hot spots structure of the proton in the impact parameter plane.
The exclusive cross section is sensitive to the average interaction of the dipole with the proton, while the dissociative one is sensitive to the fluctuations in the dipole-proton interaction due to various energy dependent color field (hot spots) configurations in the proton. The model correctly describes both ALICE and H1 data points. It also predicts the maximum of the cross section at $W_{\rm \gamma p} \sim 500$~GeV and then a decrease at higher energies. Such behavior is expected when hot spots saturate the proton, but the access to higher energies will be possible in LHC Run 3 data. 
\section{Summary}
\label{summary}
The first measurement of dissociative $\rm{J/\psi}$ photoproduction in p--Pb was shown. The result is sensitive to gluon density fluctuations in proton, but current data range does not allow for any statements on saturation. The measurements of coherent $\rm{J/\psi}$ photoproduction in Pb--Pb signal large nuclear gluon shadowing effects. The first measurement of the incoherent $\rm{J/\psi}$ photoproduction in Pb--Pb shows the importance of subnucleon fluctuations, but one should remember that fluctuations impact also the coherent regime. The theoretical description of both coherent and incoherent regime is very challenging.

\end{document}